# Perfect Structure of the Special Relativity, Superluminal, Neutrino-Photon Mass and New Entangled Interaction

*Yi-Fang Chang*
*Department of Physics, Yunnan University, Kunming 650091, China*
(e-mail: yifangchang1030@hotmail.com)

**Abstract:** First, some superluminal phenomena and experiments are introduced briefly. Next, based on the basic principles of the special relativity, the Lorentz transformation (LT) with smaller velocity v<c and the general Lorentz transformation (GLT) with larger velocity $\bar{v}$ >c should be derived simultaneously by the classification of the timelike and the spacelike intervals. In deriving LT, an additional independent hypothesis has been used, thus the values of velocity are restricted absolutely, and the spacelike interval is excluded. LT and GLT are connected by the de Broglie relation $v\bar{v} = c^2$. The fundamental properties of any four-vector and the strange characteristic which these tachyons should possess are described. The various superluminal transformations are discussed. Further, we think that LT is unsuitable for photon and neutrino, the photon transformation (PT) is unified for space $x' = r + ct$ and time $t' = t + (r/c)$. It may reasonably overcome some existing difficulties, and cannot restrict that the rest mass of photon and neutrino must be zero. LT, GLT and PT together form a complete structure of the Lorentz group. Finally, we discuss that new experiments on the quantum entangled state shown some characters, for example, coherency, nonlocality, quantum teleportation and superluminal. Further, it should be a new fifth interaction, and may probably apply GLT.

## 1. Introduction

In 1981, Pearson, et al., discovered that maps of the radio structure of quasar 3C273 shown directly that it expanded with an apparent velocity 9.6 times the speed of light [1]. Other radio sources possess also superluminal expansion. These phenomena may be explained by the relativistic beaming model [2,3], etc. Since 1995 many astronomers in different countries discovered some superluminal motions [4].

Moreover, Scharnhorst researched the propagation of light in the vacuum between plates, and their impact phrased in the simplest terms consists in causing a change in the velocity of light [5]. Barton used Euler-Heisenberg interaction, and derived the Scharnhorst effect and faster than velocity of light between parallel mirrors [6]. In this case, $c' = c/n \approx (1 - \Delta n)c > c$, in which $\Delta n = -(11\pi^2/8100)(\alpha^2/m^4L^4) \approx -1.55 \times 10^{-48}(cm/L)^4$, here L is distance between plates. But, the effects are too small by many orders of magnitude to be measured. Then Steinberg and Chiao, et al., measured the superluminal group velocity in the single-photon tunneling time [7], and observed highly superluminal propagation in a medium with a gain doublet [8]. Aharonov, et al., pointed out, unstable systems as media with inverted atomic population have been shown to allow the propagation of analytic wave packets with group velocity faster than that of light, without violating causality [9]. In 2000, Mugnai, et al., observed superluminal behaviors in the propagation of localized microwaves over distances of tens of wavelengths, and all of the components in the spectral extension have the same propagation velocity $v = c/\cos\theta$ [10]. Wang, et al., measured a very large superluminal group velocity that exceeds about 310 times faster than the speed of light in a vacuum, but these light pulses do not violate causality [11]. Marangos reviewed various superluminal effects [12].

On the other hand, the group velocity of light was dramatically reduced to as slow as 17m/s in an ultracold atomic gas, and 8m/s [13-15]. Further, Ketterle, et al., obtained an extremely slow group velocity of an amplified light pulse (~1m/s) in a Bose-Einstein condensate [16].

## 2. The classification of the timelike and the spacelike intervals.

The special relativity is based on the two basic principles: I. The special relativity principle. II. The constancy of the velocity of light in the vacuum. The mathematical representation of the second principle is invariance of the interval $s^2$. From above we may derive two symmetrical types of topologically separated structures:

| |
|---|
| A. The timelike interval $s^2 = r_{mn}^2 - c^2 t_{mn}^2 < 0$; the speed defined as $\|v\| = \|r_{mn}/t_{mn}\| < c$, which is always subluminal. |
| B. The spacelike interval $s^2 = r_{mn}^2 - c^2 t_{mn}^2 > 0$; the speed defined as $\|v\| = \|r_{mn}/t_{mn}\| > c$, which is always superluminal. |
| A. We may choose an inertial frame such that $r'_{mn} = 0$ (at the same space position), and in this case $\|v\|=0$. But $t'_{mn} = 0$ cannot be obtained. |
| B. We may choose an inertial frame that $t'_{mn} = 0$ (the simultaneity), and in this case $\|\bar{v}\| = \infty$. But $r'_{mn} = 0$ cannot be obtained. |
| A. The discussion of the time dilation must be at the same space position, X=0. From there $s^2 = x^2 - c^2 t^2 = -c^2 T^2$, and $t = T/\sqrt{1-(v/c)^2}$. |
| B. The discussion of the length contraction must involve only one time instant, T=0. From then $s^2 = l_0^2 - c^2 t^2 = R^2$, and $R = l_0 \sqrt{1-(c/\bar{v})^2}$. |

The above states do not require the Lorentz transformation. The special relativity itself does not, and cannot, restrict velocity absolutely, and does not exclude the spacelike interval and the superluminal velocity [17,18]. Because the concepts of dimensions and shape of a moving object are closely related to the concept of simultaneity [19], the simultaneity always is at the spacelike interval, which has implied the superluminal possibility.

# 3. The two symmetrical structure of the special relativity and relation between LT and GLT

In a time-space ($x_0 - x_1$) plane, a universal transformation of the special relativity is:

$$x_1' = x_1 ch\varphi - x_0 sh\varphi. \qquad (1)$$

$$x_0' = x_0 ch\varphi - x_1 sh\varphi. \qquad (2)$$

When one adds an independent hypothesis: At the same space position (for example, the origin of $K'$ frame, or for rest point, etc.), namely the premise x'=0 always exists. So we have the timelike interval, here |x/t|=|v|<c. From (1) $th\varphi = x_1/x_0 = v/c$ is obtained, and then we derive the Lorentz transformation (LT):

$$x_1' = \gamma(x_1 - vt), t' = \gamma(t - vx_1/c^2), \qquad (3)$$

where $\gamma = 1/\sqrt{1-(v/c)^2}$. The various approaches of introducing the Lorentz transformation, without exception and unanimously, applied this hypothesis [19-25]! A formula $x' = \alpha(x - vt)$ holds only for rest position of $K'$ frame. Such the spacelike interval has been excluded. LT naturally has the restriction that any velocity cannot be larger than the speed of light c.

When we give up this hypothesis, then besides (3), another symmetrical system exists yes. So long as the simultaneity t'=0 holds, thus it is at the spacelike interval, here $|x/t|=|\bar{v}|>c$. From (2) $th\varphi = x_0/x_1 = c/\bar{v}$ is obtained, then we derive necessarily:

$$x_1' = \bar{\gamma}(x_1 - c^2 t/\bar{v}), t' = \bar{\gamma}(t - x_1/\bar{v}), \qquad (4)$$

where $\bar{\gamma} = 1/\sqrt{1-(c/\bar{v})^2}$. This is called the generalized Lorentz transformation (GLT), or is called the Naranan-Jiang transformation [26,27,17].

The special interesting is that Einstein himself has expounded [22]: There are equations

$$X = ax - bct, cT = act - bx. \qquad (5)$$

If we calculate the velocity of a point (for example, the origin) of $K'$ relative to K, we have permanently X=0 (at the same space position), then

$$v = x/t = bc/a. \qquad (6)$$

LT is derived. He has also postulated to take a snapshot of $K'$ from K, calculate the velocity of an instant of $K'$ relative to K, such we have permanently T=0 (the simultaneity), then

$$x/t = ac/b = \bar{v}, \qquad (7)$$

GLT is obtained. Reversely, so also is it, if we choose K relative to $K'$.

Based on the basic principles of the special relativity, one does not add any hypothesis, we necessarily derive the two symmetrical structures: (I). The same space position x=0, at the timelike interval, the subliminal v<c，they are equivalent, LT is certainly derived. (II). The simultaneity t=0, at the spacelike interval, the superluminal $\bar{v}$ >c, they are equivalent, GLT is certainly derived. The two systems are together full of the whole time-space plane, besides the light-cone of plane, both constitute to two subgroups of the proper Lorentz group $L_p$. The velocity of light is limit one in LT and GLT. The timelike interval and the spacelike interval are partitioned into two regions of topological separation by the light-cone, both cannot hold together. Therefore, the time dilation and the length contraction, which belong to the timelike and the spacelike effects respectively, cannot also hold together. According to the GLT we obtain $l' = l_0 \sqrt{1-(c/\bar{v})^2}$, too.

So long as $\bar{v} > c$, it is always at the spacelike interval, usual LT cannot hold, only GLT holds; so the imaginary quantities do not appear naturally. On the other hand, for GLT v<c is also impossible, so any imaginary quantities cannot appear too. But, for superluminal v>c, various attempts [28-30] introduced the imaginary quantities are all based on this assumption in which LT and known formulas of the special relativity still hold in this case. It is an extension of mathematics, in spite of the condition and the regions in which the laws hold. This method will get nowhere.

Perhaps, only the timelike interval has the physical meaning. But it must add new hypothesis, otherwise the special relativity will obtain necessarily a symmetrical structure of the spacelike interval. The relation $v\bar{v} = c^2$ is derived from (6) (7) and in the references [26,27,17]. Assumption $v\bar{v} = c^2$ always holds, then $v/c = c/\bar{v}, \gamma$ and $\bar{\gamma}$, LT and GLT are only different expression, both may interchange completely, and are equivalent each other. Hence $l' = l_0 \sqrt{1-(c/\bar{v})^2} = l_0 \sqrt{1-(v/c)^2}$. Such they constitute a logical closed system, the proper Lorentz group $L_p$. We cannot but marvel with admiration that the special relativity is really a very perfect theory! If one does not escape from Einstein two basic principles, it seems to be able only to expand to this one step. Based on this scheme, an attempt which breaks essentially through the special relativity is difficult, even may be hopeless.

In a $x_0 - x_1$ plane, LT and GLT are the same equilateral hyperbola, but this region of a timelike extends to a spacelike, namely, includes the whole time-space plane besides a light-cone.

Notably, the de Broglie relation $u\bar{v} = c^2$ breaks through absolute restriction of velocity in LT at first, and is points out a possibility of the phase velocity $\bar{v} > c$ in the microphysics. Because the group velocity u is particle-velocity v, the superluminal $\bar{v}$ seems to correspond to the phase velocity, and GLT is transformation of the phase velocity among various inertial frames.

## 4. Any four-vector and tachyons

According to the above statements, any four-vector possesses following fundamental characteristics:

| Four vector | (x; ct) | $(\gamma v/c; \gamma)$ | (P; E/c) | $(j; c\rho)$ | $(A; \varphi)$ | $(k; \omega/c)$ |
|---|---|---|---|---|---|---|
| It is usually timelike vector | x<ct | v/c<1 | P<E/c=mc | $j = \rho v < c\rho$ | A<$\varphi$ | k<$\omega$/c |

| This is spacelike vector | x>ct | v/c>1 | P=mv>E/c | $j = \rho v > c\rho$ | A>$\varphi$ | k>$\omega$/c |

| Four vector | $(w_\alpha = \frac{\gamma}{c^2}\frac{d(\gamma v)}{dt}; w_0 = \frac{\gamma}{c}\frac{d\gamma}{dt})$ | | $(f\gamma/c; f\gamma v/c^2)$ | (dk; d$\omega$/c) |
|---|---|---|---|---|
| This is timelike vector | $w_\alpha < w_0$ | | 1<v/c | dk<d$\omega$/c |
| It is usually spacelike vector | $w_\alpha > w_0$ | | 1>v/c | dk>d$\omega$/c |

Let v=x/t and A=(v$\varphi$)/c, these conditions appeared usual forms can all belong to v<c. Reversely, v>c. Because $\omega$/k=v is phase velocity, (k; $\omega$/c) is usually a timelike vector with c<v. While d$\omega$/dk=u is group velocity, so (dk; d$\omega$/c) is usually a spacelike vector with c>u.

In a upper row, only x,v,p,j,A,k; $w_\alpha$, f, dk can be zero, and then LT is derived. In a lower row, only t, E, $\rho, \varphi, \omega; w_0, fv, d\omega$ can be zero, then GLT is derived. Mariwalla [31] let E=0, so GLT of the four-vector (p; E/c) was derived.

For any four-vector ($\vec{A}; A_0$), its LT is

$$A_1' = \gamma(A_1 - vA_0/c), A_0' = \gamma(A_0 - vA_1/c), \qquad (8)$$

and GLT is

$$A_1' = \bar{\gamma}(A_1 - cA_0/\bar{v}), A_0' = \bar{\gamma}(A_0 - cA_1/\bar{v}). \qquad (9)$$

Both possess the most perfect symmetrical form. Only $A_1, A_0$ interchange each other between $A_1$ and $A_0$ representations, and LT (8) and GLT (9) also interchange from v/c to c/$\bar{v}$.

Although according to the relation $v\bar{v} = c^2$, LT and GLT can interchange each other. But, if this type of tachyons in the pure special relativity really exists, since their quantities which may be zero are different, they will exhibit necessarily some new strange properties (for instance, P$\neq$0 but E=0; j$\neq$0 but $\rho$=0; A$\neq$0 but $\varphi$=0; fv$\neq$0 but f=0, etc.). Therefore, the search of tachyon is very meaning.

Though the special relativity does not exclude absolutely that certain velocity may be larger than velocity of light, but from this one may not perhaps definitely obtain the essential superluminal particles. Namely, the special relativity only describes the subliminal particles more completely, while a statement of the superluminal will await new theories, perhaps.

## 5. Various superluminal theories

The superluminal transformations of the special relativity have mainly three types as following:

A. Feinberg, Bilaniuk, et al., imaginarization method (iLT) [28-30]. Its difficulty is discussed

briefly before.

B. Naranan [26] and Jiang [27] transformation (GLT). It is based completely on the two basic principles of the special relativity.

C. Recami and Mignani transformation [32-34]. Recami and Mignani, et al., assumed $s^2 = \pm s'^2$. When v>c, $s^2 = -s'^2$, then the difference of the superluminal transformation only belongs to that a factor $\gamma$ is turned into $1/\sqrt{(v/c)^2 - 1}$. So difficulty of the imaginary quantities is overcame, and subliminal and superluminal are unified by $\gamma = 1/\sqrt{|1-(v/c)^2|}$. But, in this case, the subliminal property on the imaginary time and real space will interchange one another in superluminal, i.e., $x^2 + (ict)^2 = (ix')^2 + (ct')^2$, y and z will also be imaginary in the four dimensional time-space. Namely, the property of Murkowski time-space has larger change. Though this possibility cannot be excluded, it already included an extension of the Lorentz group and theory. Moreover, Yaccarini stated different viewpoint [35]: the linear transformation cannot derive interchange each other, and so is it surely in Lorentz group. The original Lorentz group keeps only invariance of interval $s^2$, i.e., $s^2 = s'^2$ and $s = \pm s'$. But new group is extended to $s^2 = \pm s'^2$. For superluminal case, $s^2 = -s'^2$ and $s = \pm is'$ include the real and imaginary properties interchanged each other. It is a group of all rotations in Minkowski space [32,33], and is also the simplest extension. Notably, GLT may be extended similarly yet. When $\bar{v}$<c, $\bar{\gamma}$ is turned into $1/\sqrt{(c/\bar{v})^2 - 1}$, and $\bar{v}$>c and $\bar{v}$<c are unified by $\bar{\gamma} = 1/\sqrt{|1-(c/\bar{v})^2|}$.

In this method, the timelike interval and the spacelike interval interchange each other. Of course, the superluminal is also possible. But the transversal velocity is still imaginary. Therefore, Ramanujam, et al., [36] separated $s^2$ into a transversal part $\Delta T^2 = (\Delta y)^2 + (\Delta z)^2$ and a longitudinal part $\Delta L^2 = (\Delta x)^2 - (c\Delta t)^2$. When v>c and let $\Delta L^2 = -\Delta L'^2, \Delta T^2 = \Delta T'^2$, then the x-t transformation is the same with Recami result, and the transversal velocity is real. But, if $\Delta s^2 = \Delta L^2 + \Delta T^2, \Delta s'^2 = \Delta L'^2 + \Delta T'^2$, the time-space characteristic is the same in before and after, so $s^2 \neq s'^2$. Of course, in this case we should suppose $\Delta s'^2 = -\Delta L'^2 + \Delta T'^2$, then $s^2 = s'^2$. So three dimensional real space will become one imaginary (x), and two real (y, z) space. Hence the characteristic of x and y, z are different. It is not already the Minkowski time-space. Therefore, this is more dissatisfactory. Certainly, if LT and the special relativity are extended, this result and imaginarization will be still possible.

## 6. Photon transformation and photon-neutrino mass

Recently, one believes that neutrino should possess rest mass, and three kinds of neutrino

$\nu_e, \nu_\mu, \nu_\tau$ should have three different rest masses. It is researched from various experiments and theories [37-41].

The rest mass of photon is discussed [42]. Assume that photon possesses rest mass, and if whose upper limit is determined by the uncertainty relation, the mass will be $m_0 \leq \hbar/Tc^2 \sim 10^{-66}$ g, in which $T \sim 5 \times 10^{17}$ sec is the cosmos age. This may obtain following results:

1. R.Leavitt obtained the gravitation $F \propto r^{-(2+\varepsilon)}, |\varepsilon| \leq 5.7 \times 10^{-17}, m_0 \leq 7.3 \times 10^{-48}$ g in 1983. 2. The phase velocity $u_p = c/\sqrt{1-(mc/\omega)^2}$ and the group velocity $u_g = c\sqrt{1-(mc/\omega)^2}$ will be different for electromagnetic wave in vacuum. 3. The electromagnetic wave has longitudinal polarization. 4. The constancy principle of the velocity of light may not establish. In 2002, Prokopec, et al., proposed that photon possesses mass $10^{-11}$ g from inflation at early cosmos [43].

It is well known, in usual theory this must suppose that the rest masses of all particles (for instance, photon and neutrino), whose moving velocities are the velocity of light, should be absolutely zero. Such it can be avoided that their moving masses are not infinity. But this is only a theoretical requirement. Therefore, in the last years the scientists try to measure photon-mass by the experiments, but results obtain only its upper limit. If we once discover that the rest masses of photon or neutrino are incompletely zero, whether or not the special relativity will be given up? Is it indeed the theoretical request inevitably? At least, could this supposition avoid completely these contradictories?

Based on the special relativity, according to the viewpoint of the length contraction or the time dilation in the motion of $\mu$ particle, one must recognize that when photon (including neutrino in after statement) propagates in the vacuum, the distance will contract to zero, or the time will slow to stop, namely the light will be the action at a distance. Certainly, this is absurd.

In fact, all of these difficulties originated to an implicated supposition: Photon, which is regarded as particular object, is also suitable to the Lorentz transformation and various corresponding inferences. However, this is not theoretical necessary result, and is an idealized extension arbitrarily. The special relativity derives necessarily the two symmetrical aspects: In LT $t' = \gamma(t - vx_1/c^2)$, $\gamma = 1/\sqrt{1-(v/c)^2}$. In GLT $t' = \bar{\gamma}(t - x_1/\bar{v})$, $\bar{\gamma} = 1/\sqrt{1-(c/\bar{v})^2}$. When v=c or $\bar{v}$=c, so $\gamma = \bar{\gamma} = \infty$ is meaningless, and they cannot get a definition of simultaneity, $t_n = t_m + (r_{mn}/c)$. This contradiction cannot still be overcome, even if we supposed that the rest mass of photon equals zero, or the reference, in which photon is rest, is forbidden.

We consider that the smallest correction is that photon is not suitable for LT and GLT. In the special relativity LT (v<c) in timelike interval and GLT ($\bar{v}$<c) in spacelike interval are different, so we associate easily that the transformation of light-cone (v=c), which separated two timelike and spacelike intervals, should also be different.

In the $x_0 - x_1$ plane, LT and GLT are the equilateral hyperbola $s^2 = \pm(x_1^2 - x_0^2)$, the light-cone $s^2 = 0$, which propagates photon, is an equation of straight lines $x_1 = \pm x_0 = \pm ct_0$, or is a general equation of four-dimensional time-space:

$$x_1^2 + x_2^2 + x_3^2 = c^2 t^2. \tag{10}$$

Therefore, a distance between the origins of different frames K and $K'$ is x, so the photon transformation (PT) is:

$$x' = x + ct, \tag{11}$$

or
$$t' = x'/c = t + (x/c). \tag{12}$$

Eq.(12) is also a definition of simultaneity. In this case the time-space transformation will be a formula. This is a special Lorentz group ($s^2 = 0$) without the Lorentz factor, and is also particular (v＝c) Galileo transformation $x' = x + vt$. But it has not the absolute time ($t' = t$), and the velocity of light c is always invariable.

On the dynamical aspect, we may only obtain result in which p＝mc still holds for photon and neutrino. At present our theory points out that the rest masses of photon and neutrino may not be zero, and their masses are possibly different. Such the before difficulties can mainly be solved. From this a series of new problems may be caused. Chen [44] studied the special relativity with a rest mass of photon, and in this case he thinks that the constancy of the velocity of light holds still.

## 7. The complete structure of the Lorentz group

For the two dimensional case, the Lorentz group is just a linear homogeneous transformation group which keeps invariant of square of distance $s^2$ between any two points at the $x_0 - x_1$ plane in Minkowski space. Therefore, it can only have the following cases:

A. $s^2 \neq 0$. They are the two symmetrical equilateral hyperbolas with the same asymptote.

1. $s^2 < 0, \therefore |x_1| < |x_0|$. It is in the timelike interval, and must be LT;

2. $s^2 > 0, \therefore |x_1| > |x_0|$. It is in the spacelike interval, and must be GLT.

B. $s^2 = 0$. It is the light-cone, which is the photon transformation (PT), in which both origins of K and $K'$ frames are equal. This is a common asymptote of LT and GLT. LT，GLT and PT constitute to three subgroups of topological separation, which are independent and do not interchange each other. They are connected by the relation $v\bar{v} = c^2$, and are together full of the whole $x_0 - x_1$ plane [17]. Therefore, we think that this Lorentz group is complete, in which any another transformation cannot again exist. Other transformations will correspond to extension of the Lorentz group, to extension of the special relativity. In a ward, only the Lorentz group and the special relativity are extended, then other types of transformations can be possibly derived.

The present theories suppose that the velocity of gravitational wave equals the velocity of electromagnetic wave, but the deflection of light shows that c is not invariant, and this change is different for various different gravitational fields. The gravitational wave should be propagation along straight line, and likes an electromagnetic wave in an electromagnetic field. In these cases, the two velocities are constant. But, the two velocities may not always be the same [45]. Therefore, some questions in relativity should be researched still.

The exact basic principles of the special relativity should be redefined as: I. The special relativity principle, which derives necessarily an invariant speed $c_h$. II. Suppose that the invariant speed $c_h$ in the theory is the speed of light in the vacuum *c*. If the second principle does not hold, for example, the superluminal motions exist, the theory will be still the extensive special relativity, in which the formulations are the same, only c is replaced by the invariant speed $c \to c_h$ [17,46]. Such local Lorentz transformations for different systems cannot derive varying speed of light. If the invariant speed $c_h$ are various invariant velocities, the diversity of space-time will correspond to many worlds. Moreover, based on the fractal dimensional matrix and mathematics, the fractal relativity is discussed, which connects with self-similarity Universe and the extensive quantum theory. Further, the fractal mathematics and physics may be developed to the complex dimension. The space dimension has been extended from real number to superreal and complex number. Combining the quaternion, etc., the high dimensional time $ict \to ic_1t_1 + jc_2t_2 + kc_3t_3$ is introduced. Such the vector and irreversibility of time are derived. Then the fractal dimensional time is obtained, and space and time possess completely symmetry. It may be constructed preliminarily that the higher dimensional, fractal, complex and supercomplex space-time theory covers all [47].

## 8. The entangled state is new fifth interaction

Bell discussed the Einstein-Podolsky-Rosen (EPR) paradox, and derived the well-known Bell inequalities [48]. Then he reconsidered the problem of hidden variables in quantum mechanics, which is not permitted by the demonstrations of von Neumann and others, but it is unreasonable [49]. A theorem of Bell proves that certain predictions of quantum mechanics are inconsistent with the entire family of local hidden-variable theories, and may apply to realizable experiments. Clauser, et al, proposed experiment on the polarization correlation of a pair of photons to test quantum mechanics and local hidden-variable theories [50]. Freedman and Clauser measured the linear polarization correlation of the photons emitted in an atomic cascade of calcium, and provide strong evidence against local hidden-variable theories, but agreement with quantum mechanics [51]. Shimony, et al., discussed the approximate measurement in quantum mechanics [52,53]. Clauser, et al., investigated experimental consequences of objective local theories [54], and the experiment of a polarization correlation anomaly [55]. Fry and Thompson measured the linear polarization correlation between the two photons, and test experimentally local hidden-variable theories [56].

Aspect, et al., realized Einstein-Podolsky-Rosen-Bohm gedankenexperiment by the measure

on the linear-polarization correlation of pairs of photons emitted in a radiative cascade of calcium and time-varying analyzers, and it agrees with the quantum mechanical predictions and the greatest violation of generalized Bell inequalities [57,58]. Ghosh and Mandel demonstrated the existence of nonclassical effects in the interference of two photons [59].

Then the entangled state evolves a great hotspot in physics. Kavassalis and Noolandi discussed a new view of entanglements in dense polymer systems, which predicts a geometrical transition from the entangled to the unentangled state in agreement with experimental data [60]. Horne, et al., discussed two-particle interferometry, which employs spatially separated, quantum mechanically entangled two-particle states [61]. Mermin discussed extreme quantum entanglement in a superposition of macroscopically distinct states [62]. Hardy investigated nonlocality for two particles without using inequalities for all entangled states except maximally entangled states such as the singlet state [63]. Goldstein provided a proof on Hardy theorem [64]. Kwiat, et al., reported new high-intensity source of polarization-entangled photon pairs with high momentum definition [65]. Strekalov, et al., reported a two-photon interference experiment that realizes a postselection-free test of Bell inequality based on energy-time entanglement [66].

Further, Bouwmeester, et al., investigated experimental quantum teleportation [67]. Pan, et al., experimentally entangled freely propagating particles that never physically interacted with one another or which have never been dynamically coupled by any other means. It demonstrates that quantum entanglement requires the entangled particles neither to come from a common source nor to have interacted in the past. Here they took two pairs of polarization entangled photons and subject one photon from each pair to a Bell-state measurement. This results in projecting the other two outgoing photons into an entangled state [68]. Pan, et al., reported experimental test of quantum nonlocality in three-photon Greenberger–Horne–Zeilinger (GHZ) entanglement, and three specific experiments, involving measurements of polarization correlations between three photons, lead to predictions for a fourth experiment, and found the fourth experiment is agreement with the quantum prediction [69]. Raimond, et al., performed manipulating quantum entanglement experiments with Rydberg atoms and microwave photons in a cavity, and investigated entanglement as a resource for the processing of quantum information, and operated a quantum gate and applied it to the generation of a complex three-particle entangled state [70]. Pan, et al., experimentally demonstrated observation of highly pure four-photon GHZ entanglement produced by parametric down-conversion and a projective measurement. Their technique can, in principle, be used to produce entanglement of arbitrarily high order or, equivalently, teleportation and entanglement swapping over multiple stages [71]. Zbinden, et al., reported an experimental test of nonlocal quantum correlation in relativistic configurations, in which entangled photons are sent via an optical fiber network to two villages near Geneva, separated by more than 10 km where they are analyzed by interferometers [72]. Stefanov, et al., investigated the quantum correlations with spacelike separated beam splitters in motion and experimental test of multisimultaneity [73]. Pan, et al., demonstrated experimental entanglement purification for general mixed states of polarization -entangled photons and arbitrary unknown states [74]. Yu, et al., discussed a test of entanglement for two-level systems via the indeterminacy relationship [75]. Zhao, et al., used two entangled photon pairs to generate a four-photon entangled state, which is then combined with a single-photon state, and reported experimental demonstration of five-photon entanglement and open-destination teleportation (for N = 3) [76].

Recently, Amico, et al., reviewed the properties of the entanglement in many-body systems

[77]. Korbicz, et al., shown structural approximations of positive maps and entanglement-breaking channels [78]. Orus discussed geometric entanglement in a one-dimensional valence-bond solid state [79]. Schmidt, et al., detected entanglement of a mechanical resonator and a qubit in the nanoelectromechanical systems [80]. Thomale, et al., investigated the entanglement gap separating low-energy in the entanglement spectrum of fractional quantum Hall states, and a new principle of adiabatic continuity [81]. Salart, et al., reported the first experiment where single-photon entanglement is purified with a simple linear-optics based protocol [82]. Chavez, et al., observed entangled polymer melt dynamics [83]. Jungnitsch, et al., provided a way to develop entanglement tests with high statistical significance [84]. Huber, et al., detected high-dimensional genuine multipartite entanglement of mixed states [85]. Sponar, et al., discussed the geometric phase in entangled systems for a single-neutron interferometer experiment [86]. Friis, et al., investigated relativistic entanglement of two massive particles [87]. Jack, et al., measured correlations between arbitrary superpositions of orbital angular momentum states generated by spontaneous parametric down-conversion, and quantified the entanglement of modes within two-dimensional orbital angular momentum state spaces [88]. Bussieres, et al., tested nonlocality over 12.4 km of underground fiber with universal time-bin qubit analyzers [89]. Mazzola, et al., investigate the dynamical relations among entanglement, mixedness, and nonlocality in a dynamical context [90]. Miao, et al., discussed universal quantum entanglement between an oscillator and continuous fields [91]. Chitambar, et al., considered multipartite-to-bipartite entanglement transformations and polynomial identity testing [92]. Carmele, et al., discussed the formation dynamics of an entangled photon pair [93].

In a word, many experiments on the quantum entangled state shown some new characteristics: 1. The coherency. 2. The nonlocality [63,69,72,73,89-91]. 3. The quantum teleportation [67,68,71, 74,76]. 4. The superluminal [72,73]. It already is widely applied, for example, quantum information [70], quantum swapping [71], quantum non-cloning and so on. Further, the quantum entangled state should be a new fifth interaction. Energy-time entangled photon pairs violate Bell inequalities by photons more than 10.9 km [72] and 12.4 km [89]. At present, some physicists proposed that their entangled distance is infinite, and even is an action at a distance. I think, it is middle-rang, and neither infinite nor very short. Its strength seems to obey neither the Newtonian long-range gravitational law nor the short-range interaction. It should be middle one.

It is a well-known conclusion that at the present there are only four interactions: gravitational and electromagnetic fields whose action distances are infinite (i.e., long-range forces), strong and weak fields whose action distances are very finite (i.e., short-range forces). The quantum entangled state as the new interaction will be a special place. The five interactions may be exhibited completely in Fig.1. It is in a middle state among the known four interactions, and seems to be fully enlightening meaning. Such the fifth interaction is very mystic, which is called spooky action at a distance by Einstein. Its characters are similar to the thought field [94], which may form a macroscopic entangled state, and realize a macroscopic immaterial teleportation [95].

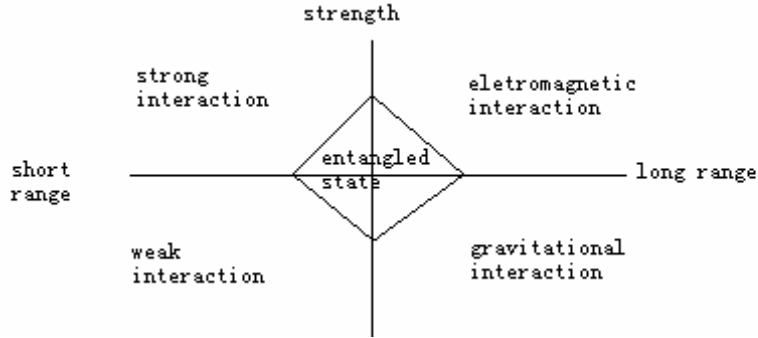

Fig.1 Relation among five interactions

At present the experiments shown that entangled photons are sent via an optical fiber network to two villages near Geneva, separated by more than 10 km. In this case the collapsed velocity of quantum state is larger than $10^7$ c [72]. The quantum correlations with spacelike are separated in motion, and experimental test is at multisimultaneity [73]. Therefore, the superluminal interaction may apply probably GLT (4) of the phase velocity, in which quantum information is a phase and corresponding velocity is possibly a phase velocity.

Myung and Kim studied the entropic force on the stretched horizon near the event horizon, and considered the similarity between the stretched horizon of black hole and the entanglement system, and defined the entropic force in the entanglement system without referring to the source mass [96]. The basis of thermodynamics is the statistics, in which a basic principle is statistical independence, and entropy is an additive quantity. If some entangled states exist among the subsystems, the second law of thermodynamics should be different. It is consistent with my conditions on possible decrease of entropy [97,98]. This should be a developed direction of the entangled state.

**References**
1.T.J.Pearson, S.C.Unwin, M.H.Cohen, et al., Nature. 290(1981)365.
2.M.J.Rees, Nature. 211(1966)468.
3.R.D.Blandford & A.Konigl, ApJ. 232(1979)34.
4.R.C.Vermeulen, Int.Astron.Union Symp. 175(1996)57.
5.K.Scharnhorst, Phys.Lett. B236(1990)354.
6.G.Barton, Phys.Lett. B237(1990)559.
7.A.M.Steiberg, P.G.Kwiat & R.Y.Chiao, Phys.Rev.Lett. 71(1993)708.
8.A.M.Steiberg, & R.Y.Chiao, Phys.Rev. A49(1994)2071.
9.Y.Aharonov, B.Reznik & A.Stern, Phys.Rev.Lett. 81(1998)2190.
10.D.Mugnai, A.Ranfagni & R.Ruggeri, Phys.Rev.Lett. 84(2000)4830.
11.L.J.Wang, A.Kuzmich & A.Dogariu, Nature. 406(2000)277.
12.J.Marangos, Nature. 406(2000)243.
13.L.V.Has, S.E.Harris, Z.Dutton & C.H.Behroozi, Nature. 397(1999)594.
14.M.M.Kash, et al., Phys.Rev.Lett. 82(1999)5229.
15.D.Budker, D.F.Kimball, S.M.Rochester, et al., Phys.Rev.Lett. 83(1999)1767.
16.S.Inouye, R.F.Low, W.Ketterle, et al., Phys.Rev.Lett. 85(2000)4225.

17. Yi-Fang Chang, New Research of Particle Physics and Relativity. Yunnan Science and Technology Press.1989. p184-209. Phys.Abst. 93(1990),No.1371.
18. Yi-Fang Chang, Galilean Electrodynamics. 18(2007),38.
19. V.Fok, Theory of Space, Time and Gravitation. New York. 1959.
20. A.Einstein, The Meaning of Relativity. Princeton. 1955.
21. L.Landau & E.Lifshitz, The Classical Theory of Field. Cambridge.1951.
22. A.Einstein, Relativity (A Popular Exposition). London. 1955.
23. P.G.Bergman, Introduction to the Theory of Relativity. New York. 1947.
24. W.Pauli, Theory of Relativity. London. 1958.
25. A.Einstein, H.A.Lorentz,•H.Minkowsky & H.Weyl,•The Principle of Relativity. London. 1923.
26. Jiang Chun-xuan, A symmetrical theory in time-space. 1972; Physics. 4(1975)119.
27. S.Naranan, Lett.Nuovo Cimento. 3(1972)623.
28. G.Feinberg, Phys.Rev. 159(1967)1089.
29. O.M.P.Bilaniuk & E.C.G.Sudarshan, Phys.Today. 22(1969)43.
30. R.G.Newton, Science. 20(1970)167.
31. K.H.Mariwalla, Amer.J.Phys. 37(1969)1281.
32. E.Recami & R.Mignani, Lett.Nuovo Cimento. 4(1972)144; 8(1973)110; 9(1974)479; Riv.Nuovo Cimento. 4(1974)209.
33. R.Mignani, E.Recami & U.Lombardo, Lett.Nuovo Cimento. 4(1972)624.
34. L.Parker, Phys.Rev. 188(1969)2287.
35. A.Yaccarini, Lett.Nuovo Cimento. 9(1974)354.
36. G.A.Ramanujam,et al., Lett.Nuovo Cimento. 6(1973)245.
37. S.E.Csorna, M.D.Mestayer, R.S.Panvini, et al., Phys.Rev. D35(1987)2747.
38. S.Abachi, P.Baringer, B.G.Bylsma, et al., Phys.Rev. D35(1987)2880.
39. E.W.Kolb, A.J.Stebbinns & M.S.Tumer, Phys.Rev. D35(1987)3598.
40. Fayyazuddin & Riazuddin, Phys.Rev. D35(1987)2201.
41. C.Q.Geng & P-Y.Xue, Phys.Rev. D35(1987)2832.
42. W.Hu, D.J.Elasenstein & M.Tegmar, Phys.Rev.Lett. 80(1998)5255.
43. T.Prokopec, O.Tornkvist & R.Woodard, Phys.Rev.Lett. 89(2002)101301.
44. I.T.Chen, Lett.Nuovo Cimento. 29(1980)518.
45. Yi-Fang Chang, Apeiron. 3(1996)30.
46. Yi-Fang Chang, arXiv. 0706.1280.
47. Yi-Fang Chang, arXiv. 0707.0136.
48. J.S.Bell, Physics. 1(1964)195.
49. J.S.Bell, Rev.Mod.Phys. 38(1966)447.
50. J.F.Clauser, M.A.Horne, A.Shimony & R.A.Holt, Phys.Rev.Lett. 23(1969)880; 24(1970)549.
51. S.J.Freedman & J.F.Clauser, Phys. Rev. Lett. 28(1972)938.
52. M.H.Fehrs & A.Shimony, Phys.Rev. D9(1974)2317.
53. A.Shimony, Phys.Rev. D9(1974)2321.
54. J.F.Clauser & M.A.Horne, Phys.Rev. D10(1974)526.
55. J.F.Clauser, Phys.Rev.Lett. 36(1976)1223.
56. E.S.Fry & R.C.Thompson, Phys.Rev.Lett. 37(1976)465.
57. A.Aspect, P.Grangier & G.Roger, Phys.Rev.Lett. 49(1982)91.
58. A.Aspect, J.Dalibard & G.Roger, Phys.Rev.Lett. 49(1982)1804.

59. R.Ghosh & L.Mandel, Phys.Rev.Lett. 59(1987)1903.
60. T.A.Kavassalis & J.Noolandi, Phys.Rev.Lett. 59(1987)2674.
61. M.A.Horne, A.Shimony & A.Zeilinger, Phys.Rev.Lett. 62(1989)2209.
62. N.D.Mermin, Phys.Rev.Lett. 65(1990)1838.
63. L.Hardy, Phys.Rev.Lett. 71(1993)1665.
64. S.Goldstein, Phys.Rev.Lett. 72(1994)1951.
65. P.G..Kwiat, K.Mattle, H.Weinfurter, A.Zeilinger, A.V.Sergienko & Y.Shih, Phys.Rev.Lett. 75(1995) 4337.
66. D.V.Strekalov, T.B.Pittman, A.V.Sergienko, Y.H.Shih & P. G. Kwiat, Phys.Rev. A54(1996)R1.
67. D.Bouwmeester, J-W.Pan, K.Mattle, M.Eibl, H.Weinfurter & A.Zeilinger, Nature. 390(1997) 575.
68. J-W.Pan, D.Bouwmeester, H.Weinfurter & A.Zeilinger, Phys.Rev.Lett. 80(1998)3891.
69. J-W.Pan, D.Bouwmeester, M.Daniell, H.Weinfurter & A.Zeilinger, Nature. 403(2000)515.
70. J.M.Raimond, M.Brune & S.Haroche, Rev.Mod.Phys. 73(2001)565.
71. J-W Pan, M.Daniell, S.Gasparoni, G.Weihs & A.Zeilinger, Phys.Rev.Lett. 86(2001)4435.
72. H.Zbinden, J.Brendel, N.Gisin & W.Tittel, Phys.Rev. A63(2001)022111.
73. A.Stefanov, H.Zbiden, N.Gisin & A.Suarez, Phys.Rev.Lett., 88(2002)120404.
74. J-W.Pan, S.Gasparoni, R.Ursin, G.Weihs & A.Zeilinger, Nature. 423(2003)417.
75. S.Yu, J-W.Pan, Z-B.Chen & Y-D.Zhang. Phys.Rev.Lett. 91(2003)217903.
76. Z.Zhao, Y-A.Chen, A.-N.Zhang, T.Yang, H.J.Briegel & J-W.Pan, Nature. 430(2004)54.
77. L.Amico, R.Fazio, A.Osterloh & V.Vedral, Rev.Mod.Phys. 80(2008)517.
78. J.K.Korbicz, M.L.Almeida, J.Bae, M.Lewenstein & A.Ac¨ˑᵃ, Phys.Rev. A78(2008)062105.
79. R.Orus, Phys.Rev. A78(2008)062332.
80. T.L.Schmidt, K.B?rkje, C.Bruder & B.Trauzettel, Phys.Rev.Lett. 104(2010)177205.
81. R.Thomale, A.Sterdyniak, N.Regnault & B.A.Bernevig, Phys.Rev.Lett. 104(2010)180502.
82. D.Salart, O.Landry, N.Sangouard, et al., Phys.Rev.Lett. 104(2010)180504.
83. F.V.Chavez & K.Saalwachter, Phys.Rev.Lett. 104(2010)198305.
84. B.Jungnitsch, S.Niekamp, M.Kleinmann, O.G¨¹hne, .Lu, W-B.Gao, Y-A.Chen, Z-B.Chen & J-W.Pan, Phys.Rev.Lett. 104(2010)210401.
85. M.Huber, F.Mintert, A.Gabriel & B.C.Hiesmayr, Phys.Rev.Lett. 104(2010)210501.
86. S.Sponar, J.Klepp, R.Loidl, et al., Phys.Rev. A81(2010)042113.
87. N.Friis, R.A.Bertlmann, M.Huber & B.C.Hiesmayr, Phys.Rev. A81(2010)042114.
88. B.Jack, A.M.Yao, J.Leach, et al., Phys.Rev. A81(2010)043844.
89. F.Bussieres, J.A.Slater, J.Jin, N.Godbout & W.Tittel, Phys.Rev. A81(2010)052106.
90. L.Mazzola, B.Bellomo, R.L.Franco & G.Compagno, Phys.Rev. A81(2010)052116
91. H.Miao, S.Danilishin & Y.Chen, Phys.Rev. A81(2010)052307.
92. E.Chitambar, R.Duan & Y.Shi, Phys.Rev. A81(2010)052310.
93. A.Carmele, F.Milde, M-R.Dachner, et al., Phys.Rev. B81(2010)195319.
94. Yi-Fang Chang, J.Rel.Psych.Res. 27(2004)126.
95. Yi-Fang Chang, J.Rel.Psych.Res. 27(2004)190.
96. Y.S.Myung & Y-W.Kim, Phys.Rev. D81(2010)105012.
97. Yi-Fang Chang, Apeiron, 4(1997)97.
98. Yi-Fang Chang, Entropy, 7(2005)190.